\documentclass{emulateapj}

\newcommand{\citepeg}[1]{\citep[{e.g.,}][]{#1}}
\def\Swift{{\textit{Swift}}\,}

\shorttitle{Radio Constraints on Dark GRB Hosts}
\shortauthors{Perley \& Perley}

\begin{document}

\title{Radio Constraints on Heavily-Obscured Star-Formation within Dark Gamma-Ray Burst Host Galaxies}

\def\cit{1}
\def\hubble{2}
\def\nrao{3}
\def\mail{*}

\author{D.~A.~Perley\altaffilmark{\cit,\hubble,\mail} and R.~A.~Perley\altaffilmark{\nrao}}

\altaffiltext{\cit}{Department of Astronomy, California Institute of Technology,
MC 249-17,
1200 East California Blvd,
Pasadena CA 91125, USA}
\altaffiltext{\hubble}{Hubble Fellow}
\altaffiltext{\nrao}{National Radio Astronomy Observatory, P.O. Box O, Socorro, NM, 87801}
\altaffiltext{\mail}{e-mail: dperley@astro.caltech.edu .}

\slugcomment{Submitted to ApJ 2013-05-13; accepted 2013-10-23}

\begin{abstract}
Highly dust-obscured starbursting galaxies (submillimeter galaxies and their ilk) represent the most extreme sites of star-formation in the distant universe and contribute significantly to overall cosmic star-formation beyond $z>1.5$.   Some stars formed in these environments may also explode as GRBs and contribute to the population of ``dark'' bursts.   Here we present VLA wideband radio-continuum observations of 15 heavily dust-obscured \Swift GRBs to search for radio synchrotron emission associated with intense star-formation in their host galaxies.  Most of these targets (11) are not detected.  Of the remaining four objects, one detection is marginal and for two others we cannot yet rule out the contribution of a long-lived radio afterglow.  The final detection is secure, but indicates a star-formation rate roughly consistent with the dust-corrected UV-inferred value.  Most galaxies hosting obscured GRBs are therefore not forming stars at extreme rates, and the amount of optical extinction seen along a GRB afterglow sightline does not clearly correlate with the likelihood that the host has a sufficiently high star-formation rate to be radio-detectable.  While some submillimeter galaxies do readily produce GRBs, these GRBs are often not heavily obscured---suggesting that the outer (modestly obscured) parts of these galaxies overproduce GRBs and the inner (heavily obscured) parts underproduce GRBs relative to their respective contributions to star-formation, hinting at strong chemical or IMF gradients within these systems.
\end{abstract}

\section{Introduction}
\label{sec:intro}

Long-duration gamma-ray bursts (GRBs; we exclude the physically-distinct separate class of short-duration gamma-ray bursts) are highly collimated and extremely luminous relativistic explosions produced during the core-collapse of massive stars.
These explosions are accompanied by extremely luminous multiwavelength afterglows \citep{Meszaros+1997,Sari+1998,vanParadijs+2000} that pinpoint the locations of their host galaxies and often reveal their redshifts as well (e.g. \citealt{Prochaska+2007b,Prochaska+2007a,DElia+2011}).   By virtue of their massive-stellar origin, the GRB rate must also be closely tied with that of overall cosmic star-formation.   For these reasons, the study of GRBs and their hosts has aroused significant interest over the past fifteen years for its potential to address greater questions of galaxy evolution and cosmic history \citep{Krumholz+1998,Totani+1999,RamirezRuiz+2002,LloydRonning+2002,Firmani+2004}.

While all GRBs appear to occur in star-forming galaxies \citepeg{Savaglio+2009}, it is not clear whether all types of star-forming galaxies can produce GRBs.   Most known GRB hosts tend to be blue, irregular, low-mass, metal-poor, and nearly dust-free; very few are spirals, have a large population of older stars, or are significantly dust-obscured \citepeg{Bloom+2002,Fruchter+2006,LeFloch+2006,Wolf+2007,Modjaz+2008,CastroCeron+2010,Levesque+2010a,Graham+2013}.  These trends have been interpreted as evidence that high-metallicity environments produce GRBs much less frequently or even not at all, an effect that would complicate the use of GRBs as a high-$z$ star-formation tracer.  Alternatively, variation in the initial mass function (IMF) or initial close-binary fraction might also result in a GRB population that seems to prefer certain types of star-forming galaxies while avoiding others.  While GRBs undoubtedly provide a wealth of information about the high-$z$ universe, placing this information into context will require a better empirical and physical understanding of how a galaxy's internal properties affect its ability to produce GRBs.

While most GRBs are only mildly obscured \citep{Jakobsson+2004,Kann+2006,Greiner+2011}, a significant minority encounter a large amount of dust along the line of sight within their host, making their optical afterglows difficult or impossible to detect.  This type of event is commonly known as a ``dark'' gamma-ray burst \citep{Groot+1998}.\footnote{In practice, the definition of a ``dark'' GRB is complex---the presence of dust extinction is only one of a large number of factors affecting the brightness of a GRB optical afterglow, and events are not followed up uniformly.  A significantly expanded discussion of the various definitions of darkness can be found in \cite{Perley+2013a}.}  Although the host galaxies of these events do tend to be more representative of the overall star-forming galaxy population then unobscured GRBs \citep{Kruehler+2011,Perley+2013a}, GRBs as a whole still appear to be underabundant in massive, reddened host galaxies compared to what would be expected for a perfect star-formation tracer \citep{Perley+2013a}.  

\begin{deluxetable*}{lllllll}  
\tabletypesize{\small}
\tablecaption{Summary of Previous Submm/Radio Host Detections}
\tablecolumns{7}
\tablehead{
\colhead{GRB} &
\colhead{Redshift\tablenotemark{a}} & 
\colhead{OA?\tablenotemark{b}} &
\colhead{Submm $F_\nu$\tablenotemark{c}} &
\colhead{Radio $F_\nu$\tablenotemark{d}} &
\colhead{Radio freq.} &
\colhead{References\tablenotemark{e}}
\\
\colhead{} &
\colhead{} &
\colhead{} &
\colhead{(mJy)} &
\colhead{($\mu$Jy)} &
\colhead{(GHz)} &
\colhead{}
}
\startdata
980425\tablenotemark{f} &  0.0085 & Y  &               &$420 \pm 50$ & 4.8 & M09  \\
031203\tablenotemark{f} &  0.105  & Y  &               &$216 \pm 50$ & 5.5 & S10   \\
\hline
000210                  &  0.8452 & N  &$3.05 \pm 0.76$&$ 18 \pm  9$ & 1.4 & T04, B03 \\
980703                  &  0.967  & Y  &               &$ 68 \pm  6$ & 1.4 & B01 \\ 
000911                  &  1.0585 & Y  &$2.31 \pm 0.91$&$          $ &     & B03 \\ 
021211                  &  1.006  & Y  &               &$330 \pm 31$ & 1.4 & M08   \\
000418                  &  1.1185 & Y  &$3.15 \pm 0.90$&$ 59 \pm 15$ & 1.4 & B03 \\
010222                  &  1.478  & Y  &$3.74 \pm 0.53$&$ 23 \pm  8$ & 4.9 & T04, B03 \\
000301C                 &  2.034  & Y  &               &$ 18 \pm  7$ & 8.5 & B03  \\
000926                  &  2.066  & Y  &               &$ 23 \pm  9$ & 8.5 & B03  \\
080607                  &  3.038  & Y  &$0.31 \pm 0.09$&             &     & W12  \\
\hline
120804A\tablenotemark{g}&$\sim$1.3& Y  &               &$ 43 \pm  4$ & 4.9 & B13 \\ 
\enddata
\label{tab:prevfluxes}
\tablenotetext{a}{\  References:  \cite{Prochaska+2004,Piro+2002,Djorgovski+1998,Price+2002,GCN1785,Bloom+2003,Jha+2001,Jensen+2001,Fynbo+2001,Berger+2013}}
\tablenotetext{b}{\  Whether or not an optical afterglow was detected for this GRB.}
\tablenotetext{c}{\  At 850$\mu$m.  The highest-confidence reported detection is given.}
\tablenotetext{d}{\  At the given frequency.  The highest-confidence reported detection (at any radio frequency) is given.}
\tablenotetext{e}{\  References for submillimeter and radio flux density measurements.  M09 =  \cite{Michalowski+2009}; S10 = \cite{Stanway+2010}; T04 = \cite{Tanvir+2004}; B03 = \cite{Berger+2003}; B01 = \cite{Berger+2001};  M08 = \cite{Michalowski+2008}; W12 = \cite{Wang+2012}; B13 = \cite{Berger+2013}.}
\tablenotetext{f}{\ Low-redshift GRBs.  Below $z\lesssim0.2$ radio observations are sensitive to ordinary (modest-SFR, optically-thin) star-forming galaxies.}
\tablenotetext{g}{\  GRB120804A is a \Swift-era event and has a duration of $T_{90} = 0.8$ sec, consistent with a short-duration burst.}
\end{deluxetable*}

Nevertheless, the mere existence of highly dust-obscured events points toward another avenue by which to investigate the relation between GRBs and overall cosmic star-formation \citep{Djorgovski+2001}.   A significant (if not necessarily dominant) amount of star formation beyond $z>1.0-1.5$ occurs within extremely luminous galaxies with star-formation rates several hundred times that of the Milky Way, almost all of which occurs behind a dust screen that is completely optically thick at optical and NIR wavelengths \citep{Smail+2004,Chapman+2005,Dye+2008,Michalowski+2010,Wardlow+2011,Yin+2012}.  These galaxies---known as submillimeter galaxies (SMGs; \citealt{Blain+2002})---appear unassuming at UV/optical/NIR wavelengths, revealing their true nature only at long wavelengths (submillimeter/radio) where the dust screen becomes transparent.

The extreme conditions in these galaxies make them important laboratories for testing whether the GRB progenitor can form in the most extreme environments. For example, submillimeter galaxies have high specific star-formation rates (sSFRs; \citealt{Daddi+2007}) but also are expected to be fairly metal-enriched \citep{Nagao+2012}.  The GRB host population appears overabundant in high-sSFR galaxies but underabundant in high-metallicity galaxies compared to expectation for a uniform star-formation tracer, so whether or not GRBs form frequently in submillimeter galaxies represents an important test regarding which of these two factors is more closely associated with GRB production.  There also have been suggestions that submillimeter galaxies (or their possible $z=0$ end-products, elliptical galaxies) exhibit an unusual IMF \citepeg{Baugh+2005,Conroy+2012}.  Determining how often these systems host GRBs (if at all) is therefore of significant interest for understanding and applying GRBs as cosmological probes.

It would be reasonable to expect that searches for the hosts of optically-bright GRBs should yield no detections at submillimeter or radio wavelengths beyond the nearby ($z\gtrsim0.5$) universe:  observations using the previous generation of instrumentation at these wavelengths are not sensitive to ordinary galaxies at these distances and can \emph{only} detect the most extreme star-forming galaxies (SFR $\gtrsim$ 200 $M_{\odot}$ yr$^{-1}$).  Nearly all galaxies with star-formation rates this high are optically thick and do not allow significant amounts of UV/optical light to escape from their star-forming regions (i.e., UV-inferred star-formation rates almost never significantly exceed this value regardless of selection criterion); this opacity should stifle the optical afterglow as well as the stellar light.   Undaunted by these pessimistic expectations, a number of submillimeter/radio GRB host surveys were conducted along primarily optically-selected and therefore unobscured sightlines anyway---and while 
a majority of the hosts targeted were indeed not detected \citepeg{Barnard+2003,Berger+2003,Tanvir+2004,LeFloch+2006,Priddey+2006,Stanway+2010,Hatsukade+2012,Michalowski+2012b}, a few optically-bright bursts \emph{were} actually localized to submillimeter-bright hosts (at least three: 980703, 000418, and 010222: \citealt{Berger+2001,Berger+2003,Tanvir+2004}; see Table \ref{tab:prevfluxes}).
The very large total star-formation rates of these systems inferred from the submillimeter/radio observations (vastly in excess of the unobscured SFR inferred from UV/optical observations; \citealt{Michalowski+2008}) leave little doubt that the vast majority of the star-formation in these galaxies is located in regions that are completely optically thick at UV/optical wavelengths.  While two of these three galaxies showed moderate obscuration, the dust columns inferred ($A_V \sim 1$ mag in the cases of GRBs 980703 and 000418 and $A_V \sim 0.1$ mag in the case of GRB\,010222; \citealt{Kann+2006}) are far less than that required to conceal the extremely luminous starbursts inferred from the radio and submillimeter data, suggesting that these GRBs were not produced by the part of the galaxy responsible for forming most of its young stars.

While this certainly indicates that the \emph{outer} regions of submillimeter galaxies form GRBs quite readily, these results are much more ambiguous about the role of the heavily-obscured inner region:  only a few well-localized ``dark'' GRBs were known at the time these early surveys were conducted, and while one of these was found to occur within a submillimeter galaxy as well (GRB\,000210), the sample of well-localized dark GRBs was far too small (and in many cases the actual evidence of dust as the cause of the optical nondetection too uncertain) to come to any definitive conclusions about the nature of dark GRB host galaxies at long wavelengths.

The launch of the \Swift\ satellite \citep{Gehrels+2004} with its on-board X-Ray Telescope has made it possible to unambiguously identify dust-obscured GRBs and accurately determine their position, and large numbers of ``dark'' GRB host galaxies are now well-characterized at optical and NIR wavelengths \citepeg{Kruehler+2011,Perley+2013a}.  However, until recently similar large efforts have not been possible at long wavelengths due to the limited sensitivity of radio and submillimeter instrumentation.
Fortunately, observational capabilities at these wavelengths are rapidly improving.  In particular, the upgraded Karl G. Jansky Very Large Array (VLA) employs new receivers and, critically, the WIDAR digital correlator which is capable of processing up to 8 GHz of bandwidth simultaneously, an increase by a factor of 80 \citep{PerleyR+2011}.  At submillimeter wavelengths, even larger gains in the sensitivity to high-$z$ galaxies are becoming available with the completion of the Atacama Large Millimeter Array (ALMA), as shown by the recent detection of one GRB host at $z=3$ \citep{Wang+2012}.

In this paper we present VLA continuum observations of 15 heavily-obscured GRB host galaxies, seeking to constrain the fraction of ``dark'' GRBs that actually originate in luminous submillimeter galaxies (versus dusty regions of more ordinary galaxies) and therefore the abundance of GRBs in this type of system.  We summarize the target selection, observational strategy, and data reduction in \S \ref{sec:obs}.   We examine our detected systems in more detail in \S \ref{sec:hosts} to explore and examine the implications of these radio detections for the nature of those objects.   Finally, we translate these observational limits into constraints on the star-formation rates and properties of the host galaxies of our sample in \S \ref{sec:discussion}, and discuss our overall results and their implications for the environments of GRBs and their use as high-$z$ SFR tracers.

\section{Observations}
\label{sec:obs}

\subsection{Targets}

\begin{deluxetable*}{llll lll}  
\tabletypesize{\footnotesize}
\tablecaption{VLA Observations}
\tablecolumns{7}
\tablehead{
\colhead{GRB Target} &
\colhead{RA} & 
\colhead{Dec} & 
\colhead{Configuration} &
\colhead{Date} &
\colhead{Int. Time} &
\colhead{RMS noise}
\\
\colhead{} &
\colhead{} &
\colhead{} &
\colhead{} &
\colhead{} &
\colhead{(min)} &
\colhead{($\mu$Jy/beam)}
}
\startdata
051008  &  13:31:29.550 & +42:05:53.30 & B  & 2012-06-11             &  70 & 3.4 \\
051022  &  23:56:04.115 & +19:36:24.04 & AB & 2011-06-06             &  30 & 3.2 \\
        &               &              & A  & 2011-08-13             &  49 &     \\
        &               &              & B  & 2012-05-26, 2012-07-19 & 194 &     \\
060202  &  02:23:23.010 & +38:23:03.20 & B  & 2012-06-23             &  46 & 5.4 \\
060306  &  02:44:22.910 &$-$02:08:54.00 & B  & 2012-06-23             &  48 & 5.9 \\
060923A &  16:58:28.160 & +12:21:38.90 & AB & 2011-06-05             &  36 & 5.6 \\
        &               &              & A  & 2011-08-21             &  23 &     \\
070521  &  16:10:38.620 & +30:15:22.40 & AB & 2011-06-05             &  36 & 4.0 \\
        &               &              & A  & 2011-08-21             &  24 &     \\
071021  &  22:42:34.310 & +23:43:06.50 & AB & 2011-06-06             &  30 & 3.8 \\
        &               &              & A  & 2011-08-13             &  38 &     \\
080207  &  13:50:02.980 & +07:30:07.40 & A  & 2011-07-17/18          &  69 & 2.4 \\
        &               &              & B  & 2012-06-05             & 187 &     \\
080325  &  18:31:34.230 & +36:31:24.80 & AB & 2011-06-05             &  36 & 2.6 \\ 
        &               &              & A  & 2011-08-21             &  23 &     \\
        &               &              & B  & 2012-06-15, 2012-06-24 & 181 &     \\
080607  &  12:59:47.221 & +15:55:10.86 & A  & 2011-07-17/18          &  71 & 3.5 \\
081221  &  01:03:10.168 &$-$24:32:51.67 & B  & 2012-06-23             &  46 & 5.9 \\
090404  &  15:56:57.520 & +35:30:57.50 & AB & 2011-06-05             &  36 & 2.6 \\
        &               &              & A  & 2011-08-21             &  23 &     \\
        &               &              & B  & 2012-05-29, 2012-06-03 & 189 &     \\
090407  &  04:35:54.980 &$-$12:40:45.50 & A  & 2011-06-18             &  55 & 4.4 \\
090417B &  13:58:46.590 & +47:01:05.00 & B  & 2012-06-11             &  68 & 4.0 \\
090709A &  19:19:42.640 & +60:43:39.30 & AB & 2011-06-05             &  36 & 5.1 \\
\enddata 
\label{tab:observations}
\end{deluxetable*}

In this study, we exclusively examine the hosts of ``dark'' GRBs---or more precisely, the hosts of GRBs whose afterglows were moderately or heavily obscured by dust as determined by combined X-ray, optical, and (often) NIR observations.  Our attention is focused on this group for two reasons.

First, the observation of a heavily-obscured GRB immediately implies the presence of at least some obscured star-formation in its host.  Heavily obscured star-formation (that behind columns of $A_V > 3-4$ mag) cannot be traced with ordinary UV or optical techniques since the UV light that traces the young stars is absorbed almost completely, and star-formation rates inferred purely from optical/UV methods alone are inherently suspect if observations of a GRB point to the presence of stars behind optically-thick dust columns within the galaxy.

Second, there is accumulating evidence that dust-obscured GRB hosts (even ones with relatively modest dust extinction columns of $A_V \sim 1-2$ mag) originate in galaxies with significantly higher average star-formation rates than ordinary hosts.   The typical host galaxy in the dark GRB host study of Perley et al.\ (2013a) has a UV-inferred star-formation rate of $\sim$50 $M_\odot$/yr and several may exceed $\sim$200 $M_\odot$/yr, which would make some of these systems feasible to detect with the VLA even without a substantial additional optically-thick component.  Moderately to heavily dust-obscured GRBs ($A_V > 1$ mag) also tend to be hosted in significantly more massive galaxies, which are much more likely to be extremely high-SFR submillimeter sources \citep{Michalowski+2008}.

The target list is a subsample of the 23 galaxies presented in \cite{Perley+2013a}, which were selected among all \Swift events between 2005--2009 with clear evidence for an afterglow extinction of at least $A_V > 1$ mag.  Nineteen of those galaxies are located above $\delta > -25^{\circ}$ and accessible to the VLA; of these we selected 15 for observations.  This (modest) down-selection was chosen based on a combination of redshift and the (UV-inferred) star-formation rate of the host, in order to further increase the prospects for detection.  Hosts at $\delta > -25^{\circ}$ presented in \cite{Perley+2013a} but not observed here are those of GRBs 060319, 061222A, 070306, and 060814.  The first three of these have low (UV) star-formation rates and high redshifts (SFR $<$ 10 $M_\odot$ yr$^{-1}$ and $z > 1$) similar to more typical GRB host galaxies at these redshifts and seemed less promising initial targets than the other galaxies\footnote{Of course, these remaining systems are also interesting in a different way, in that the detection of an obscured GRB from an unobscured galaxy directly indicates a very heterogeneous host; their blue colors and low UV SFRs are also closer analogs of the pre-\Swift submillimeter-detected hosts.  Nevertheless, for the reasons previously outlined, for this study we elected to focus on the population of more massive, higher UV-SFR hosts, leaving the remaining events for future work.}.  The host of GRB\,060814 (which has a very high UV SFR) was omitted from the sample on the basis of the apparent superposition of the host with a foreground galaxy at the time of the proposal, although subsequent higher-resolution observations successfully resolved the two components \citep{Hjorth+2012,Perley+2013a}.

\subsection{VLA Observations}

All of our observations were carried out with the upgraded VLA between May 2011 and July 2012.  We observed using the C-band (4--8 GHz) receivers, configured to position the two central frequencies at 4.715 GHz and 5.739 GHz with 1.024 GHz of bandwidth each, providing effectively continuous coverage across 2.048 GHz of bandwith centered at 5.227 GHz.  Observations were taken with 2~second (2011) or 4~second (2012) averaging and interleaved with observations of a nearby phase calibrator approximately every five minutes.

The first set observations were carried out during the summer of 2011, mostly in A configuration (resolution of 0.45\arcsec) but with some observations conducted during the reconfiguration of the array from B to A, with most antennas in their A-configuration positions but a few on the east and west arms still in B configuration.   Ten sources were observed in total during this period, with typical on-source times of approximately one hour producing typical RMS sensitivities of 5--9 $\mu$Jy.

A second set of observations were taken the following year in the B configuration (resolution 1.4\arcsec).  We re-observed four sources for which our reduction of the A-configuration data showed weak detections to significantly greater depth (an additional three hours per source), which confirmed three of these detections.  In addition, we observed five sources which we had not previously imaged in A configuration for about one hour each.

Data reduction was carried out using the Astronomical Image Processing System (AIPS).   Radio frequency interference was minor in all observations, and generally removed by clipping outlier visibilities above a minimum flux density threshold (in all cases this step removed less than 10\% of the data).  In most cases we elected not to observe a primary flux calibrator, and to instead calibrate using the switched power system, which injects a calibrated signal pattern into the data (\citealt{PerleyR2010} and work in prep.)  Comparisons of calibrations based on this system versus a standard flux-calibration procedure using 3C286 and 3C48 were found to be consistent (to within 5\%), and we expect the flux calibration not to be a significant source of error in our observations.

All observations were carried out at least two years (and more typically 4--5 years) after the GRB occurred, when the radio afterglow for typical GRBs has faded well below detectability, even to the upgraded VLA \citep{Chandra+2012}.  However, the very brightest GRBs do remain detectable for several years (e.g., Figure 25 of that work), which must be considered when assessing our putative detections. 

All observations are summarized in Table \ref{tab:observations}.

\begin{figure*}
\centerline{
\includegraphics[scale=0.57,angle=0]{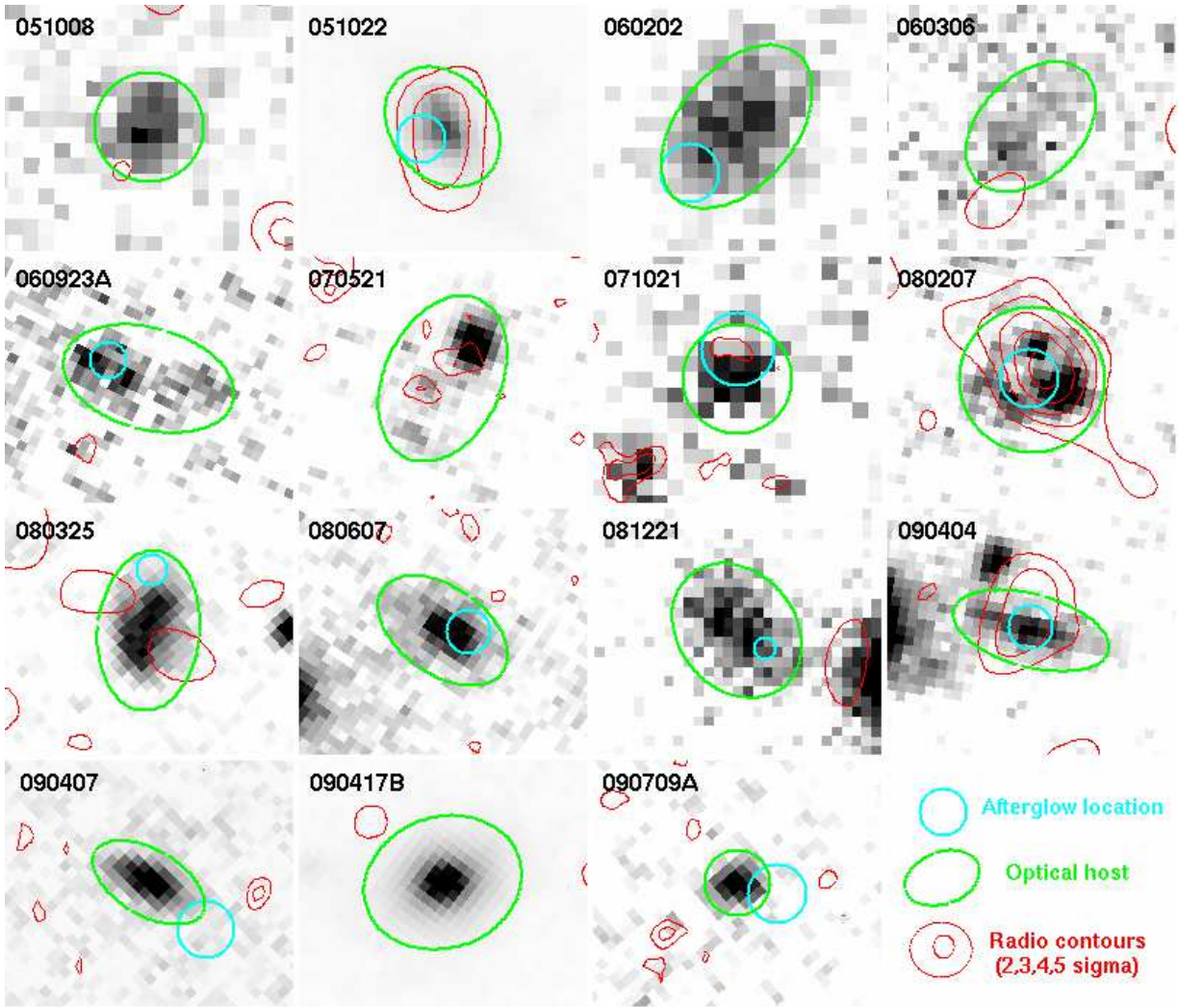}} 
\caption{Radio flux density contours (2$\sigma$ plus increments of 1$\sigma$) from VLA C-band observations of 15 dust-obscured GRB host galaxies (convolved with the default restoring beam), superimposed on optical or NIR imaging of the hosts from \cite{Perley+2013a} and uncertainty circles showing available sub-arcsecond afterglow positions (where available).  Observations of 060923A, 070521, 071021, and 080607, and 090407 were taken in A-configuration (and/or a nonstandard A/B configuration hybrid) and have a resolution of approximately 0.45\arcsec\ FWHM; the remaining observations were taken mostly in B-configuration and have a resolution of approximately 1.4\arcsec.  The 1$\sigma$ sensitivity ranges from 2.5$\mu$Jy to 6.0$\mu$Jy.  Emission at the host positions of GRBs 051008, 080207, and 090404 is detected at $>3.5\sigma$; emission at GRB\,070521 is detected only marginally.}
\label{fig:images}
\end{figure*}

\subsection{Photometry}

As our images have resolution comparable to (and often finer than) the positional uncertainty of the GRB position and the intrinsic size of the host galaxy whose flux we are attempting to measure, some care must be taken in calculation of flux densities (or upper limits) for these objects.  We provide these measurements by two different procedures.

First, we determine a \emph{point-source} flux density (or limit) on any emission underlying the position of the GRB.  This is calculated by measuring the flux density value of the brightest cell in the synthesized map within 0.4\arcsec\ of the afterglow location (or within the optical disk of the host galaxy if no precise afterglow position was available, using the images presented in \citealt{Perley+2013a}).  Physically, this corresponds to the flux (or limiting flux) associated with any ``compact'' (on the spatial scale of the resolution of the map) star-forming region that may have produced the GRB, or on any late-time afterglow emission.  The confidence of the detections are evaluated by calculating the peak flux density repeatedly within 1000 equivalent-sized regions drawn randomly from blank regions across the full synthesized map.

More importantly, we also desire a measurement (or limit) of the \emph{integrated} flux density of our hosts, corresponding to the flux emitted by all star-formation within the galaxy.  To calculate this, we further convolve the image using a Gaussian convolution kernel with FWHM set to the semimajor axis of the galaxy as measured in the optical/NIR images (see Figure \ref{fig:images}).  The flux density is determined by choosing the value of the maximum cell within that aperture, and its uncertainty estimated by repeating this exercise on random regions across the map.

The flux densities of the targets, presented using both interpretations, are presented in Table \ref{tab:fluxes}.  For most sources the estimates provide very similar results, since most of the data were taken in B configuration at resolution ($\sim$1.4\arcsec) similar to the actual angular extent of the host galaxies, so the effect of the convolution is negligible.  However, for the A-configuration-only data in particular (0.45\arcsec\ original resolution), the difference is significant, and caution should be used in interpreting the result depending on the desired goals.

\begin{deluxetable*}{lllrlllrlll}  
\tabletypesize{\small}
\tablecaption{Host Galaxy 5.23 GHz Radio Photometry}
\tablecolumns{10}
\tablehead{
\colhead{} &
\colhead{} & 
\multicolumn{4}{c}{Point-source} &
\multicolumn{4}{c}{Integrated}
\\
\colhead{GRB} &
\colhead{Beam size\tablenotemark{a}} &
\colhead{Aperture\tablenotemark{b}} & 
\multicolumn{2}{c}{$F_\nu$\tablenotemark{c}} &
\colhead{Conf.\tablenotemark{d}} &
\colhead{2$\sigma$ Limit\tablenotemark{e}} &
\multicolumn{2}{c}{$F_\nu$\tablenotemark{c}} &
\colhead{Conf.\tablenotemark{d}} &
\colhead{2$\sigma$ Limit\tablenotemark{e}}
\\
\colhead{} &
\colhead{$\arcsec$} &
\colhead{$\arcsec$} &
\multicolumn{2}{c}{($\mu$Jy)} &
\colhead{} &
\colhead{($\mu$Jy)} &
\multicolumn{2}{c}{($\mu$Jy)} &
\colhead{} &
\colhead{($\mu$Jy)} 
}
\startdata
051008  & 1.54$\times$1.34 & 0.75  &$  7.1$&$\pm$ 3.3& 0.86 &13.7 &$ 5.0$&$ \pm$ 3.6 & 0.69  & 12.2 \\
051022  & 1.41$\times$1.40 & 0.80  &$ 12.5$&$\pm$ 3.6& 0.988&     &$13.3$&$ \pm$ 3.6 & 0.988 &  \\
060202  & 1.65$\times$1.32 & 1.0   &$  5.1$&$\pm$ 5.8& 0.25 &16.7 &$ 3.4$&$ \pm$ 6.9 & 0.29  & 17.2 \\
060306  & 1.80$\times$1.27 & 1.05  &$ 13.8$&$\pm$ 6.1& 0.84 &26.0 &$14.0$&$ \pm$ 8.6 & 0.79  & 31.2 \\
060923A & 0.56$\times$0.45 & 1.1   &$ 13.1$&$\pm$ 4.2& 0.92 &21.5 &$25.0$&$ \pm$ 12.7& 0.90  & 50.4 \\
070521  & 0.57$\times$0.45 & 1.1   &$ 12.3$&$\pm$ 4.0& 0.87 &20.3 &$28.0$&$ \pm$ 10.3& 0.983 & 48.6 \\
071021  & 0.48$\times$0.44 & 0.75  &$ 10.9$&$\pm$ 3.7& 0.91 &18.3 &$12.0$&$ \pm$ 6.7 & 0.75  & 25.4 \\
080207  & 1.45$\times$1.24 & 1.0   &$ 15.5$&$\pm$ 2.3& 0.9999&    &$17.1$&$ \pm$ 2.5 & 0.9999&      \\
080325  & 1.60$\times$1.29 & 0.8   &$  6.8$&$\pm$ 2.7& 0.94 &12.2 &$ 6.4$&$ \pm$ 3.0 & 0.91  & 12.4 \\ 
080607  & 0.47$\times$0.43 & 0.8   &$  7.5$&$\pm$ 3.4& 0.54 &19.3 &$ 5.6$&$ \pm$ 6.1 & 0.36  & 17.8 \\
081221  & 2.67$\times$1.26 & 0.8   &$  6.4$&$\pm$ 5.4& 0.57 &17.2 &$ 6.4$&$ \pm$ 5.4 & 0.57  & 17.2 \\
090404  & 1.45$\times$1.30 & 0.8   &$ 10.3$&$\pm$ 2.4& 0.999&     &$10.9$&$ \pm$ 2.7 & 0.998 &      \\
090407  & 0.61$\times$0.40 & 0.75  &$  7.5$&$\pm$ 4.2& 0.31 &15.9 &$-0.1$&$ \pm$ 7.6 & 0.07  & 15.1 \\
090417B & 1.55$\times$1.37 & 1.0   &$  6.6$&$\pm$ 4.1& 0.62 &14.8 &$ 7.0$&$ \pm$ 4.8 & 0.73  & 16.1 \\
090709A & 0.88$\times$0.50 & 0.45  &$  9.0$&$\pm$ 5.0& 0.61 &19.0 &$ 1.2$&$ \pm$ 6.6 & 0.26  & 14.4 \\
\enddata
\label{tab:fluxes}
\tablenotetext{a}{\ Beam size of the optically-weighted image (major and minor axes).}
\tablenotetext{b}{\ Effective aperture used to convolve the image to produce the integrated measurements.}
\tablenotetext{c}{\ Maximum flux density (at 5.23 GHz) in any 0.1\arcsec\ synthesized cell consistent with the position of the optical/NIR host galaxy disk.  The optimally-weighted image was used for the point-source measurements; the convolved images was used for integrated measurements.}
\tablenotetext{d}{\ Significance of the detection, based on placing a large number of apertures of identical size randomly across the image and calculating the maximum flux density in each one.}
\tablenotetext{e}{\ 95\% confidence upper limit on the host flux density.}
\end{deluxetable*}

\section{Results}
\label{sec:hosts}

The majority of our fields (11 out of 15) resulted in no detection at the location of the GRB or its host.  In three of the remaining cases, we do detect clear (but weak) emission underlying the GRB/host positions:  GRBs 051022 (4.2$\sigma$ detection), 080207 (7.0$\sigma$ detection), and 090404 (3.9$\sigma$ detection).  A fourth detection (of GRB\,070521) is marginal (2.7$\sigma$ in the convolved image and a peak 3.1$\sigma$ in the unconvolved map, although the significance in the latter case is low because of the large area searched over in relation to the beam size).

Before examining the implications of our detections, it is essential to interpret their origin.  While the goal of our survey was to search for radio emission associated with intense, dust-enshrouded star-formation, alternative possibilities must also be considered:  detection of foreground/background sources, detection of AGN activity from the host, or detection of the GRB afterglow.

\subsection{AGN, Afterglow, or Host Galaxy?}

The probability of a given point on the sky intersecting an unassociated foreground or background radio source is very low: even at the $\mu$Jy level the radio sky is mostly empty, with a source density (based on inspection of our own fields) of only 2 sources stronger than 10 $\mu$Jy per square arcminute.  The probability of chance detection of an unassociated source at this level centered within $r=$1\arcsec\ of a random position on the sky is therefore approximately $P_{\rm chance} \sim r^2/(\pi \rho_{f>10\mu {\rm Jy}}^2) \sim 10^{-3}$.  For a sample of 15 positions, the probability of one or more chance alignments within the sample is still low ($P_{\rm chance} = 1.5 \times 10^{-2}$), and the chance of observing three essentially negligible ($<1\times10^{-7}$).  Our detected sources therefore in all cases almost certainly are associated with the GRB or its host galaxy in some way.

An AGN origin is also unlikely.  Statistically, most unresolved faint radio sources are star-forming galaxies \citep{Kimball+2011,Condon2012}; galaxy central black holes either tend to be much brighter (radio-loud) or are not detectable at all.  The chance of a given GRB occurring within a galaxy whose AGN radio flux density happens to land within the relatively narrow range expected for a star-forming galaxy is low:  based on Figure 11 of \citealt{Condon2012}, a detected cosmological source with a measured flux density of $<$100$\mu$Jy has about a 90\% probability to be a star-forming galaxy and not an AGN based on the relative abundance of the two populations.

While all observations were taken long after the GRB event (elapsed times range from 792--2408 days for observations of the four detected sources) and detection of an afterglow this late would be almost unprecedented (only two GRBs, 980703 and 030329, have reported detections after $>800$ days; \citealt{Chandra+2012}), the large increase in sensitivity of the VLA has rendered it possible to detect afterglows for far longer than was possible in the past.
To estimate an average statistical likelihood that an afterglow would remain bright enough at this epoch to be detected in our data, we acquired the database of all radio afterglow observations compiled by \cite{Chandra+2012} and calculated the flux density (or its limiting value) at late times by extrapolating the light curve following the last detection or upper limit as $F \propto$\ const ($t < 30$ days) or $F \propto t^{-1}$ ($t > 30$ days), a simple model that matches most events relatively well (within a factor of 2--3).    Based on this exercise, we estimate that between 10--20\% of GRB afterglows are expected to have a flux density of $>10 \mu$Jy even after 1000 days---a small minority, to be sure, but as 15 GRBs were observed during this project the detection of one or even three afterglows at this level would not be surprising.

If earlier radio data are available we can extrapolate the light curve and estimate the afterglow contribution to the late-time flux directly.  Otherwise, there are two possible means of distinguishing an afterglow origin from emission from the galaxy.  First, the spectral indices may differ: star-forming galaxies exhibit a relatively narrow range of spectral indices, between $\alpha = -1$ to $-0.5$  (\citealt{Condon1992}; we define the spectral index as $F_\nu \propto \nu^{\alpha})$.  A GRB afterglow can exhibit a much wider range of spectral indices, depending on the position of the various characteristic frequencies relative to the radio band, especially the self-absorption frequency $\nu_a$ (see, e.g., \citealt{Sari+1998} or \citealt{Granot+1999}).  In particular, a self-absorbed afterglow ($\nu < \nu_a$) will exhibit a steep spectral index of $\alpha = 2$ to $2.5$; although if the afterglow is still optically thin at late times the spectral index is less distinct;  $\alpha = 0.33$ to $-1.0$ (depending mostly on the location of synchrotron frequency $\nu_m$).  Alternatively, a GRB afterglow should always be unresolved (at VLA resolution) and be coincident (within astrometric accuracy) with earlier detections of the afterglow, while a host galaxy may be extended and/or offset from the early-time afterglow position.

With these factors in mind, we critically examine each of our four putative detections individually.

\subsection{GRB\,051022}

GRB\,051022 was observed by the VLA and the Westerbork Synthesis Radio Telescope (WSRT) on several occasions within the first week after the burst, so its late-time flux can be estimated directly.   While the radio flux is strong at early times, it declined rapidly in subsequent observations (\citealt{GCN4154,Rol+2007,Chandra+2012}); the 8.4 GHz light curve fades as approximately $t^{-1}$ even as early as 1 day.  Assuming this decay continues, the flux density should not be more than 1 $\mu$Jy at the time of our radio observation (2053--2408 days post-burst).   The more detailed broadband modeling in \citealt{Rol+2007} similarly predicts that even as soon as 30 days after the GRB the flux should have dropped below the level of our putative detection.   As a result, the source detected in our late-time observations is almost certainly not associated with the GRB afterglow and, therefore, is likely associated with the host galaxy.

The source also appears extended: we re-weighted the data within the IMAGR task to produce images of varying resolution between 1.5$\arcsec$ and 0.4$\arcsec$; the flux is seen to drop (by about 2$\sigma$) at progressively higher resolution.  This effect can also be seen as some limited north-south extension in the image even at the standard resolution.  This further supports a host-galaxy origin of the detected emission.

\subsection{GRB\,070521}

No highly significant point source appears at the location of GRB\,070521 in the synthesized map at the array's native resolution (point-source limit $F<20$ $\mu$Jy at any location consistent with the optical disk), but in the convolved image a marginal detection (98\% confidence) appears with an apparent flux density of $F = 28 \pm 10$ $\mu$Jy.  GRB\,070521 was not previously observed at long wavelengths and so its late-time afterglow flux is unconstrained, although the contribution of any point source to the detection is likely minor: the flux density value at the centroid of the radio detection is only $9 \pm 4$ $\mu$Jy in the unconvolved map.  If the source is real at all, then, it is probably also extended and therefore associated with the host galaxy.  We tentatively suggest that this source also likely represents a detection, but further observations would be necessary to unambiguously establish its reality.  We note that the location of the source is offset slightly from the brightest part of the host in the HST image, at the location of an apparent extension stretching southeast of the main disk.

\subsection{GRB\,080207}

Unfortunately, no early-time radio observations of GRB\,080207 were carried out, so we cannot directly constrain the afterglow brightness at the time of our observations.  The relatively faint and fast-fading X-ray light curve---the GRB is not detected by XRT after two days (\citealt{Evans+2009})---is not suggestive of a bright late-time radio flux, but this is far from definitive.

To attempt to constrain the spectral index of the source, we split the observations into two frequency bands: 4.828--5.468 GHz and 5.980--6.620 GHz.  The source is detected individually (at lower significance) in both of these windows separately, with a flux density of $18.9 \pm 6.6$ $\mu$Jy at 5.15 GHz and $12.7 \pm 6.3$ $\mu$Jy at 6.30 GHz.  Unfortunately, this imposes only a relatively weak constraint on the spectral index of the source ($\alpha = -2.0 \pm 3.9$) and does not rule out either model.

To evaluate the spatial extent and location of the source, we created an additional map with a (non-optimal) resolution of 0.75$\arcsec$\,$\times$\,0.75$\arcsec$ (FWHM).  The point-source flux in this map is slightly less than, but statistically consistent with, the flux in the default-resolution map, suggesting a pointlike source.  Its location ($\alpha$ = 13:50:02.96, $\delta$ = +07:30:07.4) is within the uncertainty region of the \emph{Chandra} location \citep{Hunt+2011,Svensson+2012}, and roughly halfway between the brightest optical component of the host and the fainter northern component.  As a result we cannot unambiguously distinguish the origin of this source, although if it is produced by the host galaxy itself the emission must be substantially more concentrated than the host's optical light.

The host galaxy of GRB\,080207 has been detected at 24$\mu$m with MIPS \citep{Hunt+2011,Svensson+2012}.  The 24$\mu$m flux is commonly used as an independent measure of the obscured star-formation rate of a galaxy, and the inferred value of several hundred $M_\odot$ yr$^{-1}$ \citep{Hunt+2011,Svensson+2012} is consistent with the notion of this being a heavily dust-obscured and rapidly star-forming system.  Star-formation therefore almost certainly contributes at least somewhat to the radio flux observed, although as the inferred radio SFR is even larger than this ($\sim$850 $M_\odot$ yr$^{-1}$; \S \ref{sec:sfrs}) an afterglow-dominated origin is not ruled out.

\subsection{GRB\,090404}

GRB\,090404 was also not observed in the radio at early times.   It was, however, observed and detected in the millimeter band with the Plateau de Bure Interferometer \citep{deUgartePostigo+2012}.  The reported 87 GHz flux density from this observation is 660\,$\mu$Jy at 3.4 days, which (assuming $\nu_a < \nu < \nu_m$) implies a radio flux density of $\sim$270\,$\mu$Jy at this time.  Assuming evolution similar to our generic radio light curve described earlier (flat evolution for the first 30 days, then decay as $t^{-1}$) we would expect a flux density of $\lesssim10$ $\mu$Jy at the initial observation epoch of 792 days.  While the actual afterglow contribution could easily be much less than this (in particular if $\nu < \nu_a$), again we cannot unambiguously rule out an afterglow origin to the observed detection.

Higher-resolution versions of the map produced no significant change in source flux, indicating a point-like origin (although the weak detection renders this statement not strongly constraining).  While the VLA detection is slightly north of the reported millimeter position, the offset is not significant given the uncertainties in both positions.  Similarly, attempts to subdivide the data by frequency did not produce a useful constraint on the spectral index.  As with GRB\,080207, whether this source is associated with the host or an afterglow is ambiguous.

\subsection{Star-Formation Rates}
\label{sec:sfrs}

\begin{deluxetable*}{lllll}  
\tabletypesize{\small}
\tablecaption{Host Galaxy Star-Formation Rates}
\tablecolumns{4}
\tablehead{
\colhead{GRB} &
\colhead{Redshift} & 
\colhead{UV SFR\tablenotemark{a}} &
\colhead{Radio SFR\tablenotemark{b}} &
\colhead{Point-source limit\tablenotemark{c}}
\\
\colhead{} &
\colhead{} &
\colhead{($M_\odot$ yr$^{-1}$)} &
\colhead{($M_\odot$ yr$^{-1}$)} &
\colhead{($M_\odot$ yr$^{-1}$)}
}
\startdata
051008  & $2.90^{+0.28}_{-0.15}$  &  $72^{+ 26}_{-54}$  & $<1180$     &  \\
051022  &  0.809                  &  $26^{+  7}_{-7}$   & $74\pm20 $  &  \\
060202  &  0.785                  & $5.8^{+1.1}_{-2.0}$ & $<88$       &  \\
060306  &  1.551                  & $245^{+130}_{-67}$  & $<790$      & $<560$ \\
060923A & $2.50^{+0.58}_{-0.52}$  &  $89^{+ 38}_{-30}$  & $<4140$     & $<1765$ \\
070521  & $1.70^{+1.04}_{-0.36}$  &  $40^{+ 62}_{-3}$   & $817\pm300$ & $<592$ \\
071021  &  2.452                  & $190^{+ 26}_{-20}$  & $<1828$     & $<1320$ \\
080207  &  2.086                  &  $46^{+272}_{-45}$  & $846\pm124$ &  \\
080325  &  1.78                   &  $13^{+  5}_{-4}$   & $<500$      &  \\ 
080607  &  3.038                  &  $19^{+  7}_{-5}$   & $<2070$     & $<1661$ \\
081221  &  2.26                   & $173^{+ 23}_{-30}$  & $<1030$     &  \\
090404  & $3.00^{+0.83}_{-1.82}$  &  $99^{+122}_{-99}$  & $1230\pm305$&  \\
090407  &  1.448                  &  $28^{+ 15}_{-10}$  & $<325$      &  \\
090417B &  0.345                  & $0.5^{+0.3}_{-0.3}$ & $<12$       &  \\
090709A & $1.80^{+0.46}_{-0.71}$  & $8.0^{+4.1}_{-4.1}$ & $<653$      &  \\
\enddata
\tablenotetext{a}{\  From the SED fitting procedure of \cite{Perley+2013a}, including dust-correction.}
\tablenotetext{b}{\  Calculated using the integrated flux density from Table \ref{tab:fluxes} assuming all emission originates from the host galaxy.  Uncertainties do not include the uncertainty in the photometric redshift derivations (051008, 060923A, 070521, 090404, and 090709A).}
\tablenotetext{c}{\  2$\sigma$ limit on a point-source contribution (from e.g., nuclear starburst), calculated using the point source flux in Table \ref{tab:fluxes}.  Only listed if it is significantly less than that calculated from the integrated flux; otherwise the integrated flux applies to both cases}
\label{tab:sfrs}
\end{deluxetable*}

The goal of this project was to constrain the fraction of obscured GRBs which originate from galaxies with extreme star-formation rates: ultra-luminous infrared galaxies (ULIRGs),  submillimeter galaxies, and similar classes of extreme high-$z$ star-forming objects.  To accomplish this, it is necessary to convert our flux measurements (or limits) into constraints on the host SFR.

Long-wavelength radio emission from normal galaxies is primarily generated by electrons initially accelerated by supernova remnants, but subsequently diffused into the surrounding regions of the galaxy to radiate their energy over timescales of approximately $10^7$ years via synchrotron emission.  The radio flux of a galaxy can be tied to the supernova rate, and therefore the star-formation rate \citep{Condon1992}, allowing radio observations to serve as a star-formation indicator.
Specifically, our radio SFRs are estimated using Equation 17 of \cite{Murphy+2011}\footnote{\cite{Murphy+2011} assume a \citealt{Kroupa2001} IMF, instead of the Chabrier IMF employed in \cite{Perley+2013a} to calculate the optical SFRs.  However, these two forms of the IMF are very similar and for the purposes of this paper the distinction is negligible.}:

\begin{equation}
\label{eq-sfrq}
\left(\frac{\rm SFR_{\rm 1.4GHz}}{M_{\sun}~{\rm yr^{-1}}}\right) = 6.35\times10^{-29}\left(\frac{L_{\rm 1.4GHz}}{\rm erg~s^{-1}~Hz^{-1}}\right).   
\end{equation}

Since our observations are not at 1.4 GHz, we also have to appropriately $k$-correct our observations based on an assumed spectral index $\alpha$.  As discussed above, for none of our objects do we have a robust measurement of the actual spectral index, and so we assume a canonical average value of $\alpha = -0.75$.  Incorporating this (and placing the equation in terms of the observed flux density in $\mu$Jy), we employ the following relation:

\begin{equation}
\label{eq-sfrobsq}
\left(\frac{\rm SFR_{\rm radio}}{M_\odot {\rm yr}^{-1}}\right)  = 0.072  \left(\frac{F_\nu}{\mu {\rm Jy}}\right) (1+z)^{1-\alpha} \left(\frac{d_L}{\rm Gpc}\right)^{-2} \left(\frac{\nu}{\rm GHz}\right)^{-\alpha}
\end{equation}

Using this equation, we convert our observed flux densities into measurements (or limits) on the star-formation rate of each galaxy, assuming that all of the detected emission originates from the host.  As discussed earlier, in the case of nondetections we provide separate limits (95\% confidence) on unresolved star-formation and on integrated star-formation for each galaxy. These results are given in Table \ref{tab:sfrs}.   Typical limits range from $\sim12 M_{\odot}$ yr$^{-1}$ (for GRB\,090417B at $z=0.35$) to $\sim 1000 M_\odot$ yr$^{-1}$ for targets at $z>2.5$.  Note that our uncertainty estimates do not include the impact of uncertainty in the photometric redshifts for those sources for which a spectroscopic redshift is not yet available (effects which are particularly important for GRB\,090404, which appears to be at $z\sim3$ but could be consistent with a lower redshift).

\section{Discussion}
\label{sec:discussion}

\subsection{Comparing Optical and Radio Star-Formation Rates}

The primary goal of our study is to determine the contribution of optically-thick star formation to the total.  Our radio continuum observations measure the \emph{total} (optically-thick + optically-thin) star-formation rate of each galaxy, which can then be compared to the UV/optical values which (even with dust attenuation included in the modeling) probe \emph{only} optically-thin regions.   A significant discrepancy between the two esimates (as seen in submillimeter galaxies) would indicate the presence of substantial additional optically-thick extinction.

Our sole unambiguous host-galaxy detection, GRB\,051022, has a radio star-formation rate of $74 \pm 20$ $M_\odot$ yr$^{-1}$.  This is higher, but not much higher, than our optically-derived estimates:  our UV-based estimate for this source (from SED fitting including dust absorption; \citealt{Perley+2013a}) is $26\pm7$ $M_\odot$ yr$^{-1}$.  The two values are marginally inconsistent (by 2.3 $\sigma$), hinting at the presence of modest additional, optically-thick star-formation present within this galaxy not apparent in optical obervations, even with the effects of dust reddening considered.  Given systematic uncertainties in both SFR estimators, it is not certain that this marginal excess is real, although since the GRB itself exploded in a highly optically thick region it seems likely that at least some optically thick star formation must be present.  In any case, the optically-thick star-formation rate is modest in comparison to submillimeter galaxies and previous radio-detected GRB hosts, where the optically-thick star-forming regions dominate the optically-thin ones by an order of magnitude or more.

If interpreted as host-galaxy emission (and not afterglow), the detections of GRBs 080207 and 090404 correspond to star-formation rates of $850 M_\odot$ yr$^{-1}$ and $1230 M_\odot$ yr$^{-1}$, respectively (assuming $z=3$ for GRB\,090404).  These are, indeed, greatly in excess of UV/optical estimates and would be indicative of large optically-thick star-forming regions that presumably also produced the GRB: no NIR afterglow was detected in either case.  Of course, our limits would be consistent with the optically-inferred star formation rates in the event of an afterglow origin.  Similarly, the possible detection of GRB\,070521 would imply an SFR ($\sim 850 M_\odot$ yr$^{-1}$) vastly in excess of the optical value if confirmed by further observations.

The remaining upper limits are generally well in excess of the optically-determined SFRs (by approximately a factor of 10 in most cases; see Figure \ref{fig:sfr})---consistent with the absence of a significant optically thick component, but only formally ruling out models highly dominated by such an obscured component.

We have also compared the radio star-formation rates (or limits) inferred here with dust-\emph{uncorrected} SFRs from the UV-optical SED fitting, by taking the \cite{Perley+2013a} values and re-extinguishing them using the Calzetti extinction law (the same extinction law used in the original fits).   Among our putative detections, we find ${\rm SFR}_{\rm radio}/{\rm SFR}_{\rm UV} =$ 12, 500, 2000, and 2200 for GRBs 051022, 090404 (at $z=3$), 070521, and 080207, respectively, clearly emphasizing that these are extremely dusty galaxies (typical ratios are 1--10 for ``normal'' low-$z$ galaxies, 10-500 for local LIRGs, and 1000 for Arp 220; \citealt{Howell+2010}).  Upper limits for our nondetections range from 100 (GRB\,051008) to to 2000 (GRB\,060923A).

\subsection{Implications of a Low Detection Fraction}

Regardless of interpretation for the source of the emission seen in the cases of GRBs 080207 and 090404, it is clear that we do not detect most objects in the sample.  There is no suggestion of a detection (at 95\% confidence, based on photometry of random blank locations) from either afterglow or host galaxy for 11 out of 15 fields, with typical flux density limits of $F_\nu < 15$--$20$\,$\mu$Jy (for a compact source; A-configuration-only observations have limits 2--3 times this for an extended object).

To some extent, a low radio detection fraction is to be expected: even after its upgrade, the VLA is only sensitive to the most luminous galaxies at any given redshift beyond about $z>0.2$.  To quantify these expectations, we used the observed luminosity functions for cosmological galaxy populations selected on the basis of obscured star-formation (mid-IR and far-IR field surveys) to calculate an \emph{anticipated} detection fraction, under the assumption that obscured GRBs trace obscured star-formation uniformly.

\begin{figure}
\centerline{
\includegraphics[scale=0.58,angle=0]{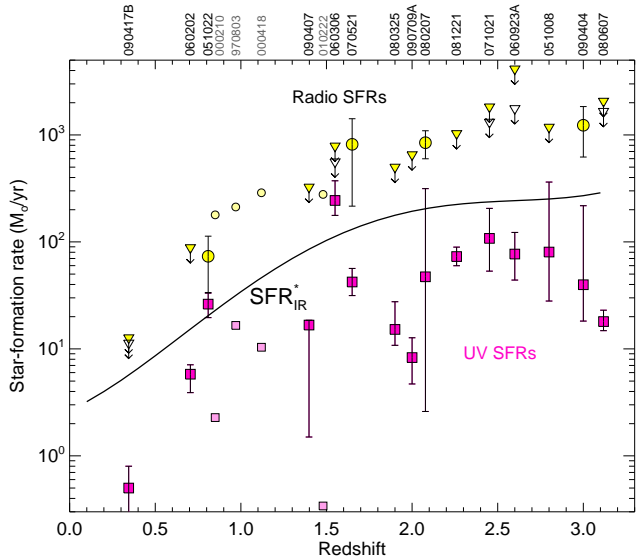}} 
\caption{Star-formation rates of targeted galaxies as a function of redshift.  Magenta squares indicate reddening-corrected UV/optical star-formation rates from the SED fitting procedure of \cite{Perley+2013a}.  Yellow points indicate our VLA measurements of the star-formation rate of these galaxies (including the contribution from optically-thick regions).   For comparison, we also show the four probable submillimeter-galaxy hosts known from pre-\Swift work (values from \citealt{Savaglio+2009} and \citealt{Michalowski+2008}) as smaller, pale symbols.  The x-axis positions of a few targets have been offset slightly from their actual redshift values for clarity.  The solid curve shows the star-formation rate of an $L_*$ galaxy (from the IR luminosity functions of \citealt{PerezGonzalez+2005} and \citealt{Lapi+2011}) as a function of redshift.}
\label{fig:sfr}
\end{figure}

Infrared luminosity functions are drawn from the \cite{Schechter1976} function fits of \citealt{PerezGonzalez+2005} (Spitzer rest-frame 12$\mu$m, for $z<1.5$) and from \citealt{Lapi+2011} (Herschel/ATLAS rest-frame 100$\mu$m, for $z>1.5$; we performed our own Schechter fits to their data with $\alpha$ fixed to $-1.2$).  We then convert the $L^*$ parameter of the 12$\mu$m or 100$\mu$m luminosity function to an equivalent SFR at each redshift using the $F_{\rm IR}$-to-SFR conversions as detailed in each of the two reference works\footnote{The conversions employed by \citealt{PerezGonzalez+2005} assume a Salpeter IMF, so we convert to Chabrier by scaling the SFR down by a correction factor of 1.6.}.  (This SFR$^*$ is plotted as the solid curve in Figure \ref{fig:sfr}.)  We then convert this SFR$^*$ to an equivalent observed 5 GHz flux density $F_{\nu, {\rm radio}}^*$ at each redshift using the \cite{Murphy+2011} relation.  The value of the Schechter parameter $\alpha$ is unaltered by the conversions (since the scaling is linear to a good approximation) and the normalization is unimportant, so this converts the mid-IR/far-IR luminosity Schechter function, $\phi_{L{\rm(IR)}}(z) \propto (L/L^*)^\alpha\,{\rm e}^{-L/L*}$ to a predicted radio flux density function, $\phi_{F}(z) \propto (F/F^*)^\alpha\,{\rm e}^{-F/F*}$.  

Galaxies with brighter radio fluxes have proportionally higher star-formation rates, so the relative fraction of obscured star-formation at a given redshift occurring in all galaxies brighter than our typical flux density limit $F_{\rm (obs)}$ is given by a normalized integral of $\phi_{F}$ weighted by the radio flux density: 
$${\rm frac}_{\rm SFR}(F > F_{\rm obs}) = C^{-1}  \int_{F_{\rm obs}}^{+\infty} F \phi_{F}\,d{\rm F} $$
($C$ is a normalization constant equal to the integral of the term on the right from $-\infty$ to $+\infty$.)

This value is redshift-dependent, though only by a factor of a few over the range of our sample: at higher-$z$ the increase in importance of highly star-forming galaxies partially compensates for flux dimming.   We find that only a small, albeit non-negligible, fraction of obscured SFR (2--5\% at $0.6 < z < 2$; more at lower redshifts and less at higher redshifts) is in radio-detectable galaxies to an average limit of $\sim15$ $\mu$Jy.  Detecting only one galaxy in the sample would be entirely consistent with a host population that traces obscured star-formation.  Detection of several (accepting the detection of 070521 as genuine and interpreting 090404 and 080207 as galaxy, not afterglow, emission) would actually indicate a population skewed in \emph{favor} of the most luminous galaxies.

\subsection{On The Relative Detection Fractions of Dark and Non-Dark Bursts}

Interestingly, non-dark GRBs are also sometimes found in very luminous hosts dominated by optically thick star-formation, even though very little unobscured star-formation in the universe occurs in such systems.
Even if we conservatively consider only those host galaxies with secure detections at both submillimeter and radio wavelengths\footnote{A number of other possible detections have been reported in radio or submillimeter wavelengths alone (\citealt{Berger+2003,Tanvir+2004}; see Table 1).  But since submillimeter measurements are often confusion-limited and radio observations are subject to afterglow contamination, it is not clear if all of these represent genuine host detections.} there are at least two highly secure, IR ultra-luminous (inferred $L_{\rm IR} \gtrsim 10^{12} L_{\odot}$) host galaxies belonging to GRBs that were not heavily obscured: GRBs 000418 and 010222.  Considering the fact that the control sample of pre-\Swift GRBs for which radio/submillimeter host searches have been conducted was only a few dozen, this corresponds to a fraction of $\sim10$\%.  While still a minority, this fraction is much larger than would be expected given that these events were optically-selected, as the fraction of \emph{optically-thin} star-formation in the universe occurring within submillimeter galaxies is almost negligible (less than 1\%).  These galaxies would therefore appear to be extremely effective (optically-thin) GRB producers given their (optically-thin) star-formation rates.

Yet, it cannot be the case that the \emph{entire} galaxy experiences an elevated GRB rate.  Submillimeter galaxies are completely dominated by optically-thick star-formation, so if the GRB rate is the same throughout this type of galaxy, then for every mildly-obscured GRB localized within a submillimeter galaxy, there should be many more dark GRBs occurring within similar systems.   Our new results show clearly that this is not the case: heavily obscured dark GRBs are already relatively uncommon ($\sim$15\% of the population) and only a few of them at most occur within bright submillimeter galaxies.  In submillimeter galaxies, therefore, \emph{only} the optically-thick outer regions seem to show a large amplification of the GRB rate relative to star-formation, providing evidence that the GRB rate relative to that of overall star-formation varies within galaxies as well as between them.

Alternative interpretations are possible, although appear unlikely.  If the gas and dust densities within submillimeter galaxies were sufficiently high the column would become Compton-thick, suppressing the X-ray afterglow completely and preventing the afterglows and host galaxies from ever being localized.  However, since the prompt gamma-rays would generally not be absorbed these events would be manifest as a large ``X-ray-dark'' population, the presence of which appears to be ruled out: almost every \Swift\ GRB ($\sim96$\%) that can be followed rapidly with the XRT is successfully detected in the X-ray band \citep{Burrows+2008}.

It is also possible that the unobscured pre-\Swift GRBs from submillimeter host galaxies actually \emph{did} occur inside highly obscured regions of their hosts, but were able to destroy the dust along the line of sight by photoevaporation and/or X-ray grain shattering \citep{Waxman+2000,Fruchter+2001,Draine+2002,Perna+2002,Perna+2003}, allowing the afterglow to emerge mostly unimpeded.  In this case, our results would be much less surprising.  However, for this to be possible nearly all the obscuring dust would have to be concentrated quite close to the GRB (dust destruction is only thought to be effective within $\sim$10 pc).  It is difficult to understand how the dust in these galaxies would be entirely concentrated within such a small distance of the massive stars and yet have such a high covering fraction (up to 99\%).  Furthermore, evidence of dramatic early-time extinction variations from rapid follow-up of GRB afterglows are so far lacking (\citealt{Oates+2009,Perley+2010}, although see also \citealt{Morgan+2013}).

We are left, then, with the likelihood that the inferred trend is real, and that optically-thin star-forming regions within submillimeter galaxies are indeed much more likely to produce GRBs than the optically-thick regions.  While the physical cause of this trend is not clear, almost any interpretation would have interesting implications for both the formation mechanism of GRBs as well as the structure of submillimeter galaxies.

The most popular interpretation for the apparent variation of the GRB rate between galaxies is that it is a metallicity effect: high-metallicity environments underproduce GRBs and low-metallicity environments overproduce them \citep{Modjaz+2008,Levesque+2010a,Graham+2013}.  SMGs and ULIRGs do indeed seem to show significant differences in metallicity between their optically-thick inner regions and optically-thin outer regions \citep{Swinbank+2004,Rupke+2008,Caputi+2008,Santini+2010,Nagao+2012}, which could explain our observations.
Nevertheless, the contribution of optically-thin regions of submillimeter galaxies to metal-poor cosmic star-formation is (like the contribution of these regions to star-formation overall) likely to be very low, so it is still not clear whether this effect alone can explain the observations or whether some other condition more specific to submillimeter galaxies---such as a non-standard IMF \citep{Baugh+2005,Conroy+2012}---might also be required to explain the observations.

These different possibilities---any of which would be quite significant---clearly demonstrate the need for continued submillimeter and radio follow-up of GRB host galaxies for understanding the environmental dependences of these enigmatic objects.  Our observations only skirt the tip of the luminosity function (typical upper limit of $3L*$)---and while pre-\Swift observations were similarly sensitive only to these depths, deeper observations might unveil more highly starbursting hosts.  Our interpretations also depend critically on observations of a relatively small number of pre-\Swift hosts conducted over a decade ago.  Very little radio and submillimeter follow-up of GRB hosts has been conducted since 2004, most of which has targeted galaxies at $z<1$---a period when optically-thick star-formation played only a minor role in the overall cosmic story.    

With the upgraded VLA, SCUBA-2, and ALMA all now available, renewed investigation of the ``dark'' side of GRB host galaxies in these dust-unbiased long-wavelength windows are now more practical than ever before, even at $z>2$ and beyond.  Future long-wavelength follow-up of the host galaxies of both obscured and ``ordinary'' GRBs will help to further illuminate the role of different types of environments in producing the GRB phenomenon.

\vskip 1cm

\acknowledgments

We thank the referee for many helpful suggestions that improved the content and readability of this manuscript.
We also acknowledge helpful comments from M. Micha{\l}owski, D.~Frail, and D.~A. Kann, and valuable discussions with N.~Tanvir.  We further thank D.~Frail for sharing the database of GRB afterglow radio light curves.
Support for this work was provided by NASA through Hubble Fellowship
grant HST-HF-51296.01-A awarded by the Space Telescope Science
Institute, which is operated by the Association of Universities for
Research in Astronomy, Inc., for NASA, under contract NAS 5-26555.
The National Radio Astronomy Observatory is a facility of the National Science Foundation operated under cooperative agreement by Associated Universities, Inc.


{\it Facilities:} \facility{VLA}


\end{document}